\documentclass[twocolumn,aps,prb]{revtex4}
\usepackage{epsf,graphicx}
\begin{document}
\title{{\bf Mechanical properties of the domains of titin in a Go-like model}}

\author{Marek Cieplak\footnote{
Correspondence to: Marek Cieplak,
Institute of Physics, Polish Academy of Sciences,
Al. Lotnik\'ow 32/46, 02-668 Warsaw, Poland;
Tel:  48-22-843-7001, Fax:  48-22-843-0926;
E-mail: {\sl mc@ifpan.edu.pl}}
}
\affiliation{ Institute of Physics, Polish Academy of Sciences,
Al. Lotnik\'ow 32/46, 02-668 Warsaw, Poland}

\author{Annalisa Pastore}
\affiliation{National Institute for Medicine Research,
Department of Molecular Structure, The Ridgeway-Mill Hill,
London NW7 1AA, UK}

\author{Trinh Xuan Hoang}
\affiliation{Institute of Physics and Electronics, Vietnamese Academy of
Science and Technology, 10 Dao Tan, Ba Dinh, Hanoi, Viet Nam}

\begin{abstract}
Comparison of properties of three domains of titin, I1, I27 and I28,
in a simple geometry-based model shows that despite a high structural
homology between their native states different domains show similar
but distinguishable mechanical properties. 
Folding properties of the separate domains are 
predicted to be diversified
which reflects sensitivity
of the kinetics to the details of native structures.
The Go-like model corresponding to the experimentally
resolved native structure 
of the I1 domain is found to provide the biggest thermodynamic
and mechanical stability
compared to the other domains studied here.  
We analyze elastic, thermodynamic and kinetic properties of several
structures corresponding to
the I28 domain as obtained through homology-based modeling.
We discuss the ability of the models of the I28 domain to
reproduce experimental results qualitatively.
A strengthening of contacts that involve hydrophobic amino acids
does not affect theoretical comparisons of the domains.
Tandem linkages of up to five identical or different domains
unravel in a serial fashion at low temperatures.
We study the nature of the intermediate state that arises
in the early stages of the serial unraveling and find it
to qualitatively agree with the results of Marszalek et al.
\end{abstract}

\maketitle

\section{INTRODUCTION}
The way proteins are able to sustain mechanical stress has recently
attracted large interest and promoted the development of new tools, both
experimental and theoretical, to study mechanical unfolding.
\cite{Carrion1999,Carrion2000,Li2000}
One model system largely
used for such studies is titin, a giant modular protein specific for
vertebrate muscle. A single-chain titin molecule forms a filament up to
1-2$\mu m$ long which connects the edge of the sarcomere, the basic unit of
the muscle fibrils, with its middle (for reviews see 
Ref.\cite{Maruyama1994,Maruyama1997,Keller1995,Trinick1996}
).  These connections provide both a molecular
ruler that determines the exact length of the sarcomere and a template for
interactions with other proteins involved in muscle ultrastructure and
regulation. 
\cite{Trinick1996}
One of the most important functions of titin
is to act as a spring which confers passive elasticity 
on sarcomers 
\cite{Horowits1989,Horowits1992,Funatsu1993,Trombitas1991,Wang1991}.

Two sequence motifs are present in the I-band, the elastic region of
titin: a long stretch rich of prolines, glutamic acids, valines and
lysines (PEVK motif) and up to (depending on the isoform) 100 copies of
tandem bead-like globular domains whose fold belongs to the immunoglobulin
(Ig) superfamily.\cite{Labeit1995} 
Elasticity is thought to
result from the interplay of these two elements acting as molecular
springs placed in series, in which most of the tension is provided by the
largely unstructured PEVK motif, whereas the Ig domains could provide an
additional contribution through a reversible unfolding mechanism.
\cite{Linke1998,Linke1996}
Most of the recent studies of
titin elasticity have concentrated on the Ig element taking the 27th 
Ig module of cardiac titin I-band (I27) as a representative model system. 
Its  three-dimensional structure has been 
determined -- 
it is an 8 strand $\beta$ sandwich with two
anti-parallel $\beta$-sheets packed against each other and held together 
by a tight hydrophobic core. \cite{Improta1996} 
The N- and C-termini point 
to opposite directions, thus making the motif particularly suitable to a
sequential assembly in a filament. Thermodynamically, I27 is highly stable
both against the thermal and the chemical unfolding.
\cite{Politou1995,Politou1996} 
The folding pathways both of the isolated I27 and of a
homopolymer constructed by tandem I27 repeats have been characterised in
detail experimentally.
\cite{Fowler2001,Carrion1999a,Carrion2000,Marszalek1999}
Molecular dynamics (MD) studies
and other simulations have provided further details into the mechanism of
unfolding suggesting that detachment of the A-strand is the earliest step
in forced unfolding. 
\cite{Best2001,Lu1999,Lu2000,Paci2000,Klimov2000}

The folding, elastic, and thermodynamic properties of 
of I27 have also been
studied within the Go model.
\cite{Cieplak2002b,Cieplak2004a,Cieplak2004b}
Go models \cite{Abe1981,Takada1999} 
are constructed based on the
knowledge of the native structure and are coarse-grained.
In their simplest version, the protein is represented by
the locations of the $C^{\alpha}$ atoms. This modelling is
geometry-based and is implemented by choosing
effective couplings between the $C^{\alpha}$s in a way
that the ground state of the system coincides with the
native conformation of the protein. This approach is less
realistic than all-atom simulations but it offers many
advantages. It allows one for studies of: a) folding and stretching within
the same model, b) tandem arrangements of many domains, c) ranges
of control parameters such as temperature, $T$, d) more realistic
pulling speeds, $v_p$, and e) differences and similarities between
various proteins within one framework. It also 
highlights  the link between native structure and the
properties of a protein. However, it is expected that
the further away from the
native structure a conformation is, the more approximate the
description becomes.

The focus of this paper is to consider two other Ig domains from the
elastic region of titin, I1 and I28, and to compare {\it in silico} 
their mechanical, kinetic, and thermodynamic properties to those 
of I27 within 
the same theoretical framework. The three domains correspond to distinct 
sequences and their level of identity is low -- it ranges between
30\% and 40\%. \cite{Fraternali1999} 

The structure of I28 is available from homology based 
modelling, \cite{Fraternali1999} 
whereas the native structure of I1 has been determined 
experimentally by x-ray crystallography.\cite{Mayans2001} 
The thermodynamic stabilities of the domains have also been measured and show 
a remarkable difference between I1 and I27 (which are very stable) 
and I28 (the most unstable).\cite{Politou1995} 
The refolding kinetics of I28 also has been found to be about
three orders of magnitude slower than that of I27.\cite{Li2000}
Despite its low thermodynamic stability, the experiments show that 
I28 is mechanically
more stable than I27.\cite{Li2000}
Here, we produce a number of homology-based models of I28 and compare
their behavior. However, in most of the paper, we focus on the model
derived in.\cite{Fraternali1999} 
We show that none of the structures generates
a full qualitative agreement with the experimental findings, at least
when analyzed within the dynamical framework provided by a simple Go-like
model.

\vspace*{0.5cm}

\section{SOURCE OF THE STRUCTURES USED}

The experimental structures of I1 and I27 have been deposited  in the
Protein Data Bank\cite{Bernstein1997} 
We shall refer to them hereon as I1-1g1c and I27-1tit. 
The I1 domain consists of 98 residues whereas I27 of 89 residues.
The other two structures have been determined by homology 
modelling\cite{Fraternali1999} 
and will be denoted as I1-model and I28-model. 
I28-model was obtained using the MODELLER program.
\cite{Sali1993} 
The I1-model is the structure of a 
mutated sequence of I1-1g1c at a single position Gly-71-Ala.
This mutation occurs in the well exposed loop region,
denoted by H in the lower left panel of Figure 1, and is not
expected to affect the thermodynamic or mechanical stability of the
domain. It is probably an inconsequential mutation that occurs
spontaneously in the system used in crystallographic studies.
The root mean
square deviation between this structure and the experimental I1-1g1c
is about 1.16 $\AA$. 
Ribbon representations of the native structures of the domains
studied here are shown in Figure 1. 

At the end of the paper, we shall discuss other 
homology-based models of the I28 domain. These are denoted by I28-A
through I28-E. I28-A and I28-B were obtained by the PSQPIR routine
in the WHATIF program. \cite{Vriend1990} 
The templates used here
were 1tit.pdb and 1tiu.pdb respectively -- the former represents
the average structure of the NMR bundle and the latter is the
first structure of the NMR bundle (this structure is the best
in terms of the internal energy).
I28-C and I28-D used the same templates but were produced by
the automatic Swissmodel webserver. \cite{Guex1997} 
Finally, I28-E was produced by the Swissmodel webserver
using the 1tlk.pdb template. The alignment, produced by clustalx, 
\cite{Thompson1997} 
was in any case straightforward since the sequences of I27
and I28 have the same length and do not require insertions/deletions. The
sequence of telokin (1tlk) can also be structurally aligned to I27
producing a unique alignment.

\begin{figure}
\epsfxsize=3.2in
\centerline{\epsffile{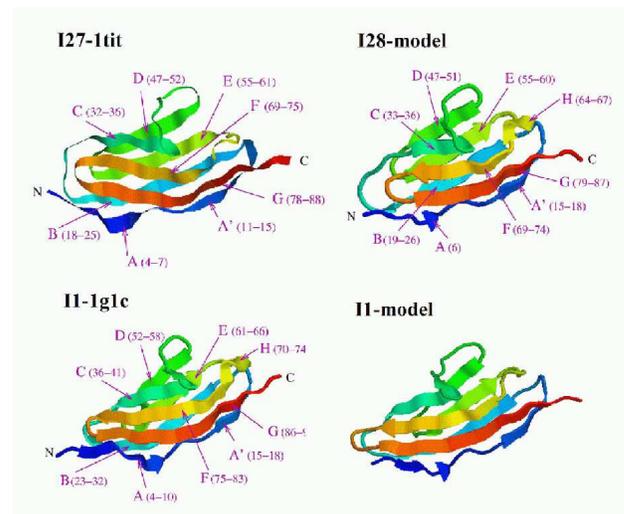}}
\caption{The structures of the domains of titin studied in this
paper. The labelling of the $\beta$-strands (symbols
A through G) is indicated, together with
the allocation of the amino acids to the structures.
The fragments corresponding to the $\alpha$-helix are denoted by H.
}
\label{fig:structures}
\end{figure}

\begin{figure}
\epsfxsize=3.2in
\centerline{\epsffile{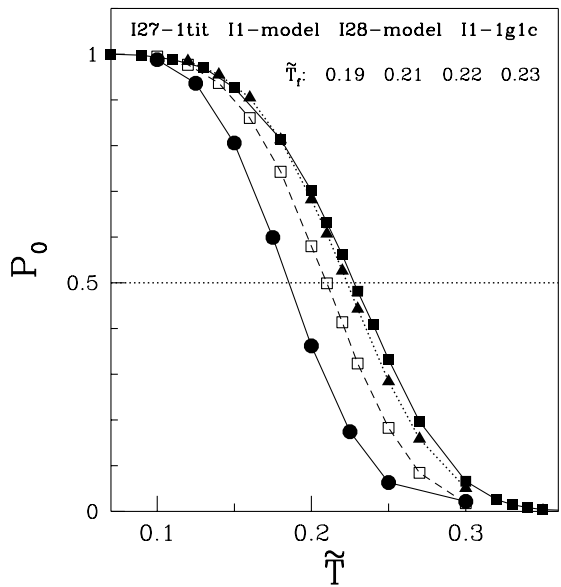}}
\caption{Probability of staying in the native structure for the
domains indicated. The lines correspond to the domains left-to-right
as listed in the order at the top of the figure.
The corresponding values of $\tilde {T}_f$ are listed below in the
same order.
}
\label{fig:stabi}
\end{figure}

\begin{figure}
\epsfxsize=3.2in
\centerline{\epsffile{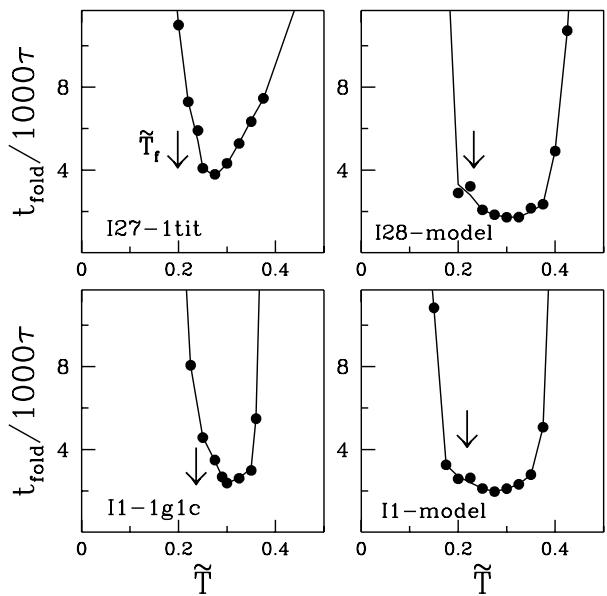}}
\caption{ Median folding times for the four systems as a function
of temperature. The arrows indicate values of the folding temperature
$\tilde{T}_f$. All of these systems are good folders in the sense that
the folding temperatures are in the region of good, if not necessarily
optimal, folding. The error bars are of the order of the data
points (the convention kept throughout the paper).
}
\label{fig:czasy}
\end{figure}

\begin{figure}
\epsfxsize=3.2in
\centerline{\epsffile{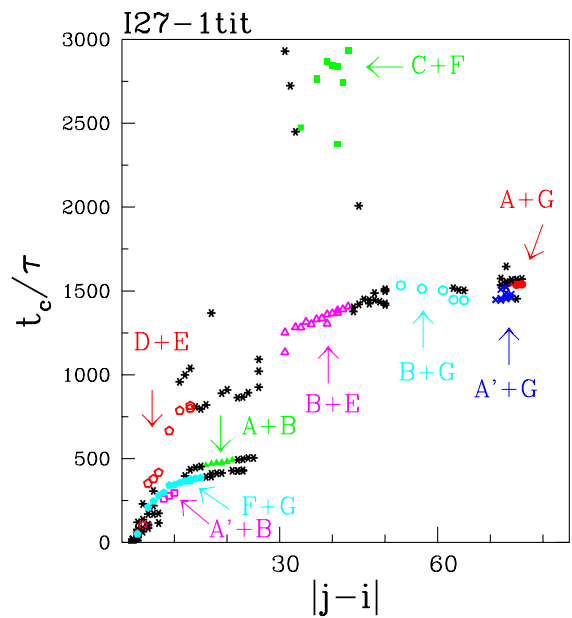}}
\caption{The scenario of folding events for  I27-1tit.
The data points show average times (the average is over 200 trajectories)
to establish specific contacts corresponding to the sequential
distance of $|j-i|$ at the temperature of the optimal folding.
The letter symbols indicate the nature of the strands that form the
contacts.}
\label{fig:scenfol}
\end{figure}

\section{MODEL AND METHOD}
We perform molecular dynamics simulations of a continuum space
Go-like model. The coarse-grained
character of our theoretical description was motivated by the desire
to deal with models that allow for studies of folding.
The details of our approach are described in 
Ref. \cite{Hoang2000,Hoang2001} 
with refinements as presented in Ref.\cite{Cieplak2003a}.
Each amino acid is represented by a point particle of mass $m$ located
at the position of the C$^{\alpha}$ atom.
The interactions  between amino acids are divided into native and 
non-native contacts. 
We follow a procedure given in 
Ref. \cite{Cieplak2002}
and determine the native contacts by considering the
all-atom native structure and by identifying those pairs
of amino acids whose atoms effectively overlap.
In the criterion of the effective overlap, the atoms are
represented by spheres with radii that are a factor of 1.24 larger
than the atomic van der Waals radii 
\cite{Tsai1999} 
to account for the softness
of the potential. The native contacts are then represented
by the Lennard-Jones potentials
$4 \epsilon [ (\sigma_{ij}/r_{ij})^{12} - (\sigma_{ij}/r_{ij})^{6}]$,
where $r_{ij}$ is the distance between 
C$^\alpha$ atoms $i$ and $j$. The length parameters $\sigma _{ij}$
are determined so that the minimum of the pair potential coincides
with the distance between C$^\alpha$ atoms in the native structure.
In order to prevent entanglements, the remaining pair-wise
interactions, i.e. the non-native contacts, correspond to a
pure repulsion. This is accomplished by taking the Lennard-Jones
potential with $\sigma_{ij}=\sigma=5$ $\AA$ and truncating it
at $2^{1/6}\sigma$.

All contacts have the same energy scale $\epsilon$. This energy
scale corresponds to between 800 and 2300 K as it represents
effectively hydrogen bond and hydrophobic interactions so the
room temperature should correspond to $\tilde{T}=k_BT/\epsilon$ of
about 0.1 -- 0.3 ($k_B$ is the Boltzmann constant). The specificity
corrections could be implemented if known reliably.
Replacing the Lennard-Jones potential in the contacts by a
10-12 interaction yields equivalent results 
both in folding \cite{Cieplak2003b} 
and in stretching.\cite{Cieplak2004b} 
An improved simulation might
involve enhancing the strength of contacts that correspond
to a disulfide bridge that is present in the I1 domain.

Neighboring C$^\alpha$ atoms are tethered by a harmonic potential
with a minimum at 3.8$\AA$ and the force constant of
$ 100 \epsilon $\AA$^{-2}$.
A four-body term that favors the native sense of the local chirality 
and thus facilitates formation of local structure in a proper way
is also kept in the Hamiltonian.\cite{Cieplak2003a}
This term vanishes for the native-like chirality and introduces an energy
penalty of order $\epsilon$ for the opposite chirality \cite{Kwiecinska}.

A Langevin thermostat with damping constant $\gamma $ is coupled
to each C$^\alpha$ to control the temperature.
For the results presented below $\gamma = 2 m/\tau$, where
$\tau =\sqrt{m \sigma^2 / \epsilon} \sim 3$ps is the characteristic time
for the Lennard-Jones potential.
This produces the overdamped dynamics
appropriate for proteins in a solvent,\cite{Cieplak2003a}
but is roughly 25 times smaller than the realistic damping from water,
\cite{Veitshans1997} 
Previous studies show that our choice speeds the kinetics up without
altering behavior, and tests with larger $\gamma$ confirm a linear
scaling of folding times with $\gamma$.
\cite{Hoang2000,Hoang2001} 
Thus the folding times reported below should be
multiplied by 25 for comparison to experiment.

The stretching protocol follows the reference
\cite{Cieplak2002a} 
and is implemented parallel to the initial
end-to-end vector of the protein and both ends of the protein
are attached to harmonic springs of spring constant $k$.
We consider the "soft" spring case of $k\;=\;0.12 \epsilon / $\AA $^2$
which corresponds to typical elastic constants of AFM cantilevers.
The outer end of one spring is held fixed at the origin, and the
outer end of the other is pulled at constant speed $v_p$
which results in a displacement $d$ away from the location
at the initial time.
We focus on $v_p = 0.005$\AA$/\tau$
which corresponds to a velocity of about $7 \times 10^{6}$ nm/s
when $\gamma = 2 m/ \tau$.
Experimental AFM velocities range from 0.3 to 10 000 nm/s
\cite{Marszalek1999,Rief1997,Fowler2002}
whereas all
atom simulations correspond to speeds which are at least six orders
of magnitude higher.\cite{Lu1999} 
This large speed used in the all-atom simulations is
believed to be one of
the reasons for the peak forces that are a factor of 10
bigger than found experimentally (another, and
probably more important, could be working
against the surface tension of the droplet of water
that surrounds the model I27 domain of titin).

The folding and stretching processes are characterized by the order in
which native contacts are formed and broken respectively.
The complication is that, at a finite $T$, a pair distance $r_{ij}$
may fluctuate around a selected cut-off value. Thus, when
discussing folding, we determine
the average time $t_c$ for each contact to form for
the first time. On the other hand, when discussing stretching,
we determine the average displacement, $d_u$,
at which a contact holds for the last time.
The presence of a contact between
amino acids $i$ and $j$ is declared when $r_{ij}$
does not exceed 1.5$\sigma _{ij}$.

\begin{figure}
\epsfxsize=3.2in
\centerline{\epsffile{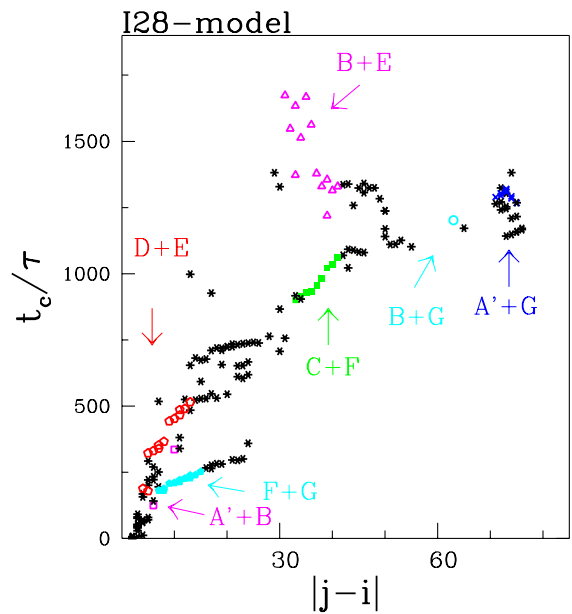}}
\caption{Same as in Figure 4 but for I28-model. }
\end{figure}

\begin{figure}
\epsfxsize=3.2in
\centerline{\epsffile{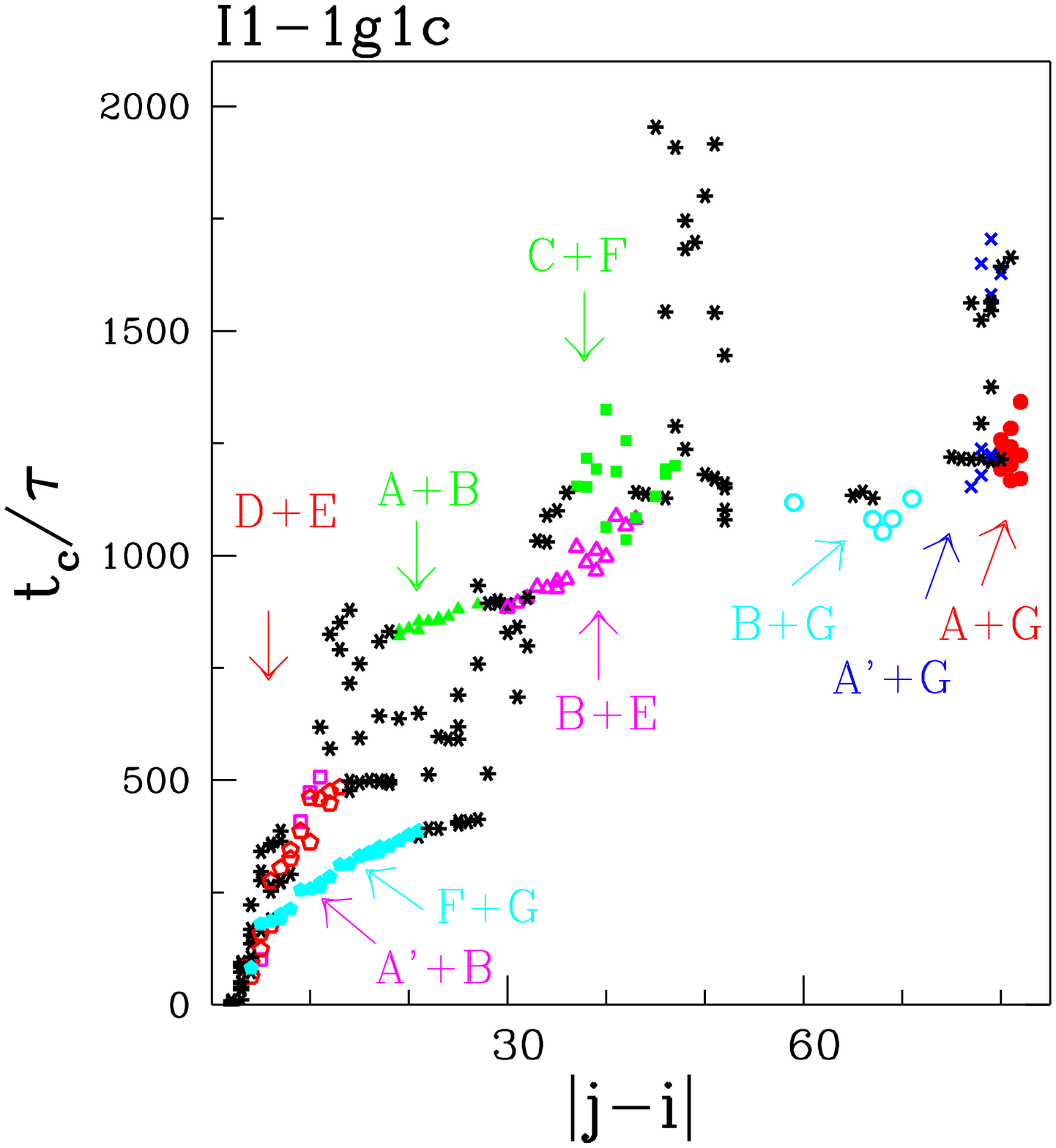}}
\caption{Same as in Figure 4 but for I1-1g1c. }
\end{figure}

\begin{figure}
\epsfxsize=3.2in
\centerline{\epsffile{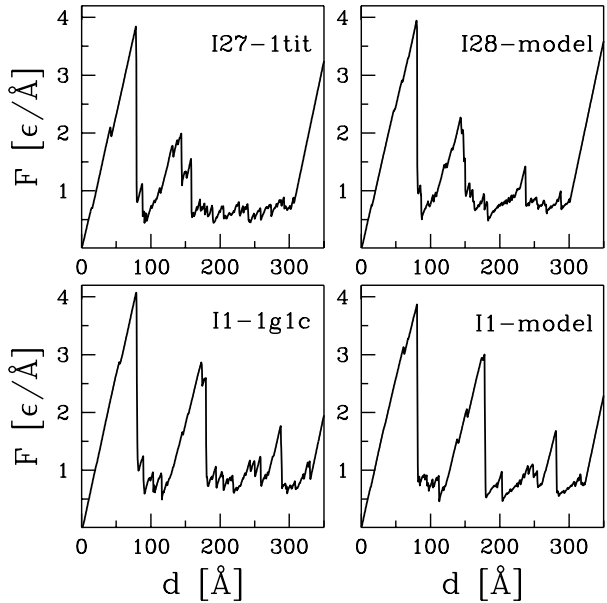}}
\caption{Force--displacement curves for the four structures
at $\tilde T=0$. }
\label{fig:pasofto}
\end{figure}

\begin{figure}
\epsfxsize=3.2in
\centerline{\epsffile{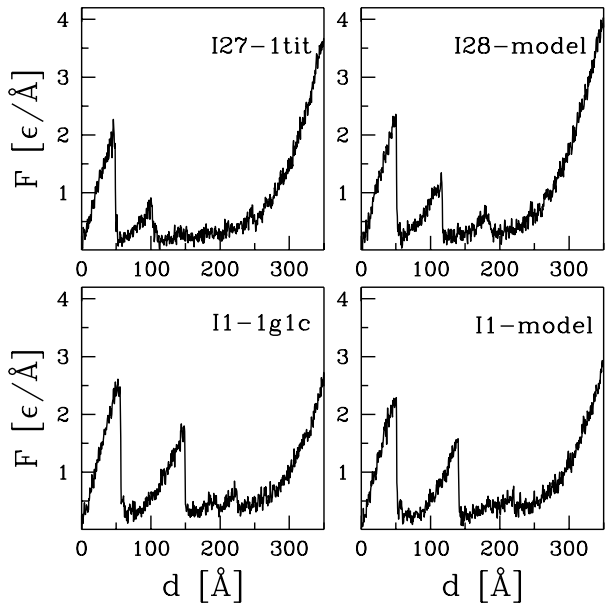}}
\caption{Same as in Fig. 4 but for $\tilde{T}$=0.3 which should
correspond to the room temperature situation.
The pulling force is averaged over 100$\tau$,
i.e. over the distance of 0.5 $\AA $,  to reduce the random noise.
}
\label{fig:pasofto3}
\end{figure}

\section{RESULTS AND DISCUSSION}

\noindent
{\bf Folding of single domains of titin}\\

Figure 2 shows the equilibrium probability, $P_0$, of staying
around the native state as a function of $\tilde{T}$.
The criterion for this is that each pair that forms a native
contact does not exceed the cut-off distance.
$P_0$ depends on $\tilde{T}$ in a sigmoidal fashion and the folding
temperature, $T_f$, corresponds to $P_0$ crossing 0.5. The values
of $\tilde{T}_f$ do not vary much: they range between 0.19 and 0.23.
The smallest value is for I27-1tit and the largest for
I1-1g1c, suggesting that domain I1, at least when in isolation, should
be more stable than I27. The value of $\tilde{T}_f$ for I1-model
is 0.21 -- it is close to I1-1g1c but clearly the two values are
not identical. I28-model appears to be more thermodynamically stable 
than I27 which is at variance to the experimental
findings.\cite{Politou1995} 

In order to characterize the folding kinetics we start the system
in an unfolded conformation and determine the median 
"first passage time", $t_{fold}$, i.e. the first time to establish
all native contacts. The $\tilde{T}$ dependence of $t_{fold}$
is shown in Figure 3. The two homology-determined structures
have a broad region of fast folding and $\tilde{T}_f$ is within
this region. On the other hand, the experimentally determined 
structures correspond to a significantly reduced width of
the region of
best folding with $\tilde{T}_f$ being just outside of this region
and on the low temperature side. The optimal values of $t_{fold}$
indicate that the structure I27-1tit is the hardest to fold to.
Despite the difference in the $\tilde{T}$ dependence between
I1-model and I1-1g1c the two systems have nearly the same
optimal folding times.

Figures 4 through 6 show that the four
systems (results for I1-model not shown) also differ in their folding
scenarios as represented by plots of $t_c$ vs. the contact
order, $|j-i|$, at the temperature of the fastest folding. 
I1-model shows the most monotonic dependence of
$t_c$ on $|j-i|$. However, its counterpart, I1-1g1c,
concludes its folding by establishing mid-range contacts.
The same happens for the remaining structures but the nature
of the last contacts is specific to the structure: it is 
the C strand joining F
in the case of I27-1tit and the B strand joining E in the case of
I28-model. Each scenario involves nearly parallel branches
of events taking place at various times for the same contact order.
The four systems studied here are clearly not equivalent
kinetically and, in particular, I1-model is not equivalent to I1-1g1c.
\\

\begin{figure}
\epsfxsize=3.2in
\centerline{\epsffile{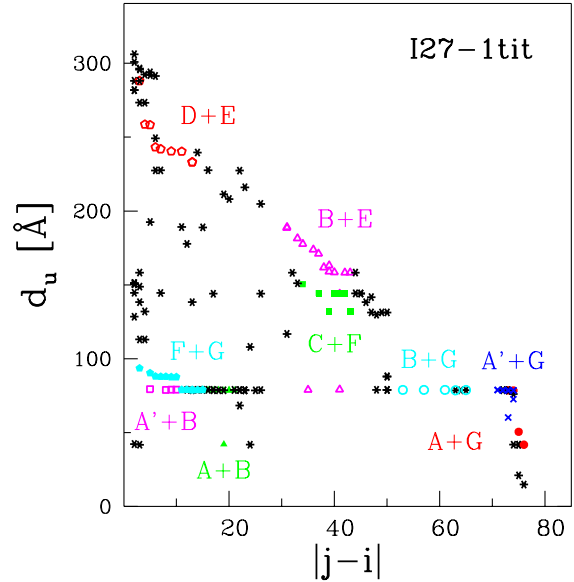}}  
\caption{The scenario of mechanical unfolding at $\tilde{T}$=0
for I27-1tit.
The data points show the last distance at which particular contacts
are considered to be still holding. The letter symbols are as in Figure 4.}
\label{fig:stretch}
\end{figure}

\begin{figure}
\epsfxsize=3.2in
\centerline{\epsffile{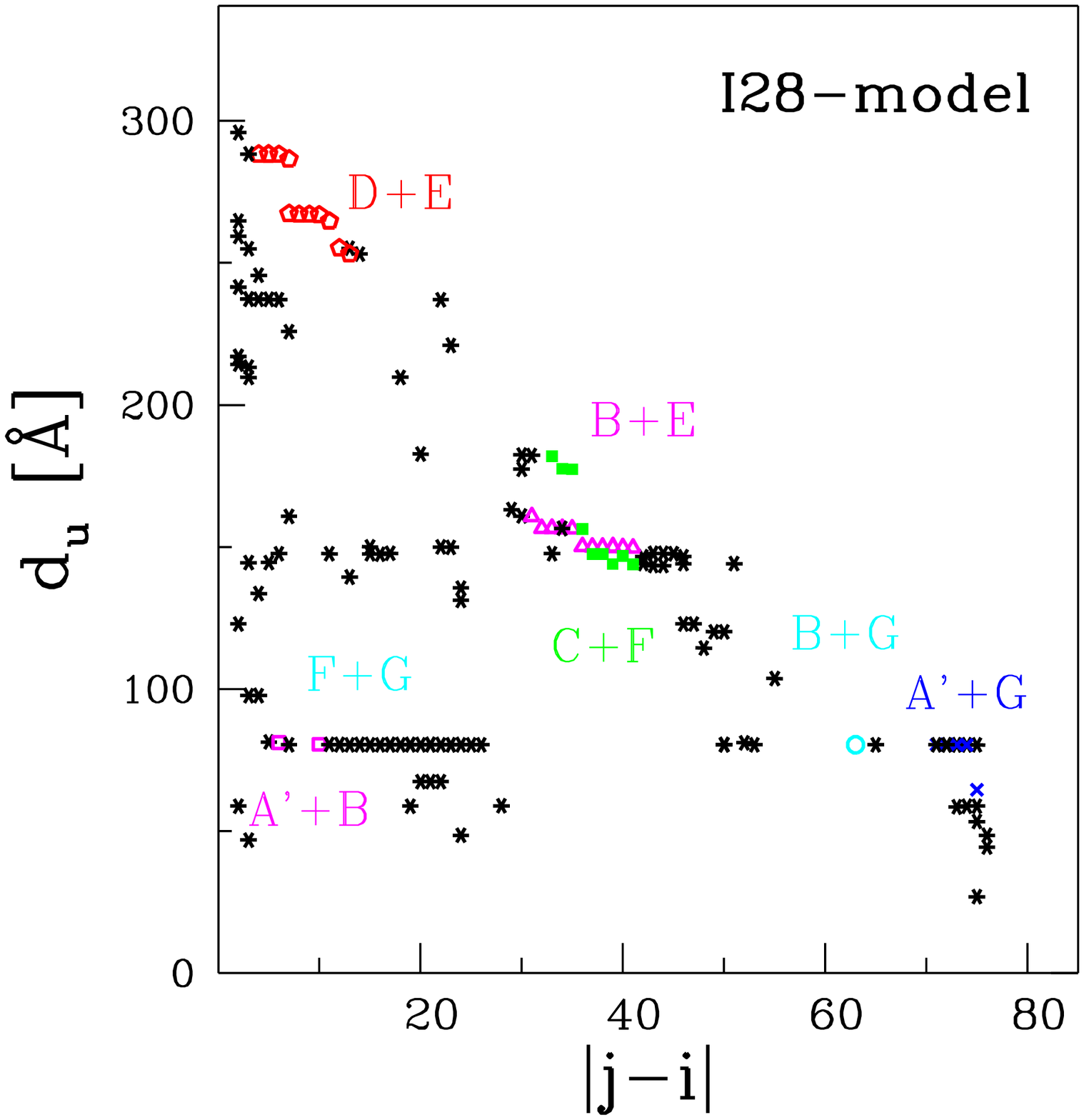}}  
\caption{Same as in Figure 9 but for I28-model.}
\end{figure}

\begin{figure}
\epsfxsize=3.2in
\centerline{\epsffile{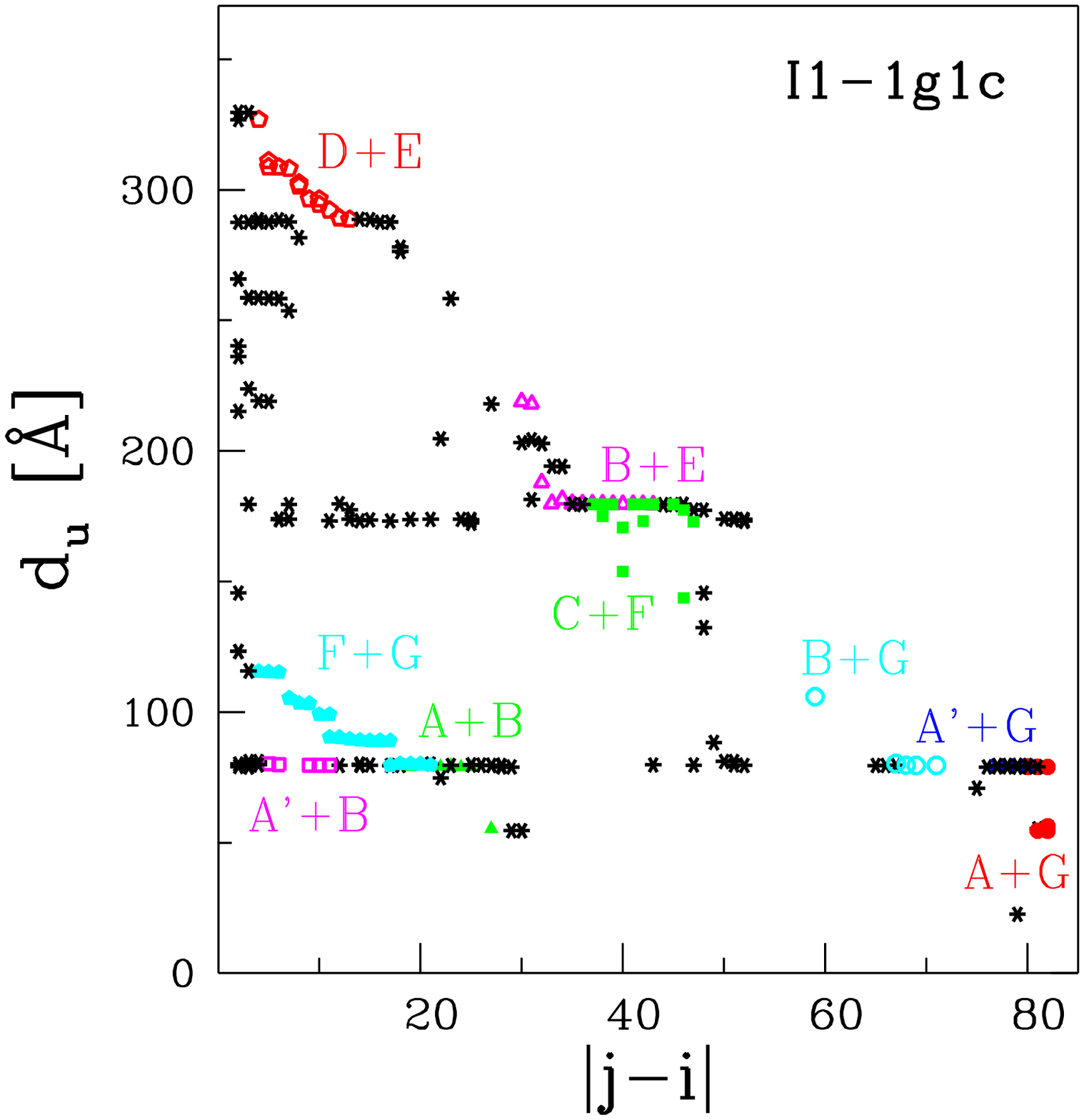}}  
\caption{Same as in Figure 9 but for I1-1g1c.}
\end{figure}

\begin{figure}
\epsfxsize=3.2in
\centerline{\epsffile{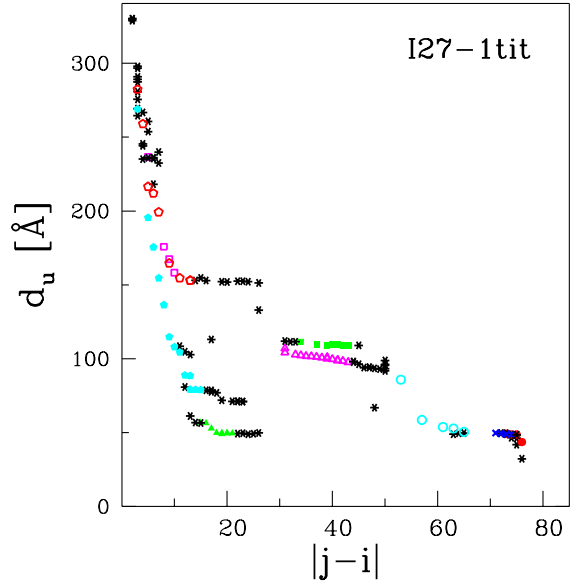}}
\caption{Same as in Fig. 9 but for $\tilde{T}$=0.3.
The data points are averaged over 20 trajectories.}
\label{fig:stretch3}
\end{figure}

\noindent
{\bf Stretching of single domains of titin}\\

The dependence of the pulling force, $F$, on the displacement
for the four structures
is shown in Figures 7 and 8 for $\tilde{T}$=0 and 0.3 respectively.
All of the four structures have quite similar $F-d$ patterns.  
The maximum force peak is large and in all cases it comes early during 
the unfolding providing the main resistance of the structures to the
pulling device.
The maximum force peak is the highest for I1-1g1c at both temperatures.
This indicates that when standing alone I1 is the most stable 
domain to sustain pulling force. The heights of the maximum forces
of other three structures are very close to each other. 
At both temperatures, I1-model appears to be mechanically less stable 
compared to I1-1g1c. 

The second maximum in the patterns shows more variations. 
The details of its shape varies between the domains
and its height is noticeably bigger for the two
versions of the I1 domain. Unlike the first maximum, the second
maximum builds up at a location which differs somewhat from structure
to structure. The third maximum is practically absent in the case of
I27-1tit but it exists, in differing shapes, in the remaining three
structures. 
The differences in the shapes and positions of the maxima
are related to different sets of contacts that are involved.
We note that, mechanically, I1-model appears to be nearly 
indistinguishable from
the experimental structure I1-1g1c.
Overall, the differences in the $F-d$ curves between the
studied domains are subtle so the domains
should be functionally equivalent in stretching.

The differences between the structures diminish as the
temperature is raised and at a sufficiently high temperature, of 
order 0.8 in this case, the $F-d$ curves switch to a
worm-like chain featureless behavior \cite{Marko1995}
in which $F$ just grows
with $d$ monotonically combined with small thermal 
fluctuations.\cite{Cieplak2004a,Cieplak2004b} 

The scenarios of the contact rupture at $\tilde{T}$=0 and 0.3
are shown in Figures 9 through 11 and 12 through 14 respectively.
Sets of data points that form nearly horizontal lines
correspond to the maxima in the $F-d$ curves. 
The nature of the first maximum is very similar for all of the structures
studied and it involves rupturing of the A'--G contacts.  The first
to rupture are the strictly terminal A--G bonds but this 
process generates no big force.
There are also varying contributions from the A'--B, A--B, B--G and
F--G contacts. The second maximum involves separating strand C from F
and a varying number of contacts between strand B and E.  The third
maximum (not present for I27-1tit) 
involves breaking of contacts between 
strands D and E.

Marszalek et al. \cite{Marszalek1999} have identified a folding
intermediate in I27 -- a hump on the $F-d$ curve that precedes
the maximum force -- as being due to the rupture of two hydrogen
bonds in A--B. This identification relies on the 
AFM technique combined with
the steered molecular molecular dynamics simulations and on 
making an amino-acidic substitution on the sixth position.
Our $\tilde{T}$=0 results suggest that most of the
A--B bonds (the filled triangles in Figures 9 and 12; in Figure 9
these symbols overlap with other and, except for one, are hard to see)
break simultaneously with those of A'--G and one (between
4 and 23) precedes the rupture of A'--G. At $\tilde{T}=0.3$, on the other
hand, the bonds of A'--G and A--B rupture almost simultaneously.
It should be noted that the coarse grained nature of the Go-like model
does not allow to identify breaking of hydrogen bonds. The hydrogen
bonds between the A and B strands arise probably at
5--24 and 6--24. Since we use  relatively large cut-off
distances for contact breaking, these contacts are still present
when the 4--23 bond is ruptured. However, the  relative
displacement of strand A with respect to strand B should 
deform the hydrogen bonds.
We conclude that our advance rupture of the
4--23 bond observed in the low temperature data
can be related to the hydrogen bond rupture
discussed in ref. \cite{Marszalek1999}.
Thus our results are in a qualitative agreement with the
previous finding. 

As the temperature is raised from $\tilde{T}$=0 to 0.3,
the contact rupturing scenarios simplify somewhat: the events
the contacts are more clearly grouped into several horizontal lines
corresponding to the maxima in the force. 
For instance, the breaking of the A--G
contacts at $\tilde{T}$=0.3 takes place almost at the same time as that 
of the A'--G bonds. 
The heights of the force peaks, however, decrease due to
thermal fluctuations.
On increasing the temperature further, the stretching scenarios
become even more simplified. 
At $\tilde{T}$=0.8, i.e. in the entropic limit, they become
strictly monotonic as a function of the contact order,
as illustrated in Figure 15. In this limit, the four systems are
strictly indistinguishable from the mechanic point of view.
\\

\begin{figure}
\epsfxsize=3.2in
\centerline{\epsffile{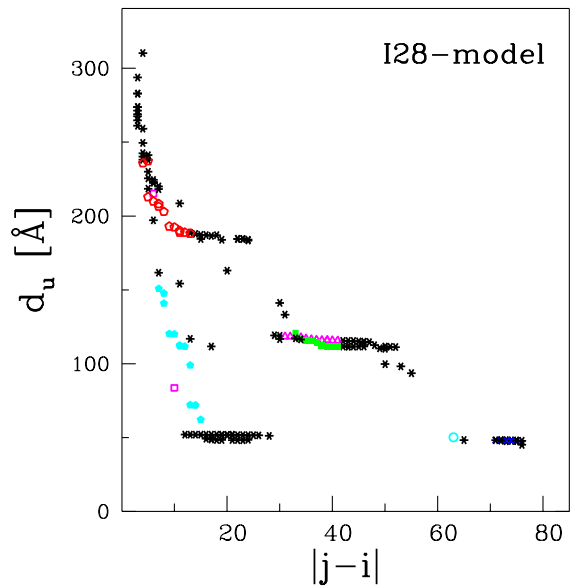}}
\caption{The scenario of stretching for I28-model 
at $\tilde{T}=0.3$. }
\end{figure}

\begin{figure}
\epsfxsize=3.2in
\centerline{\epsffile{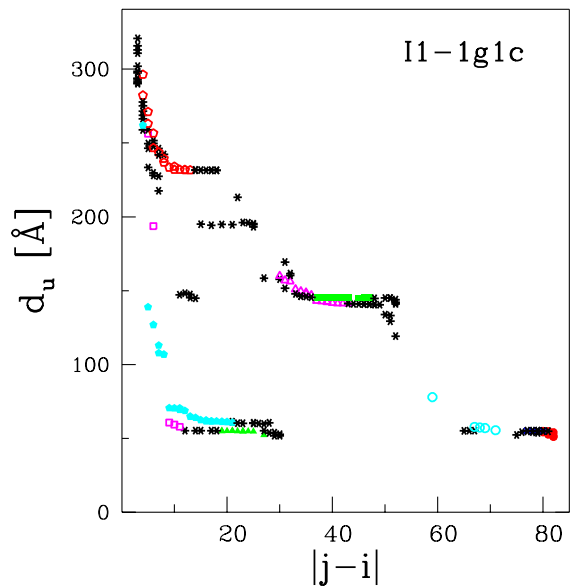}}
\caption{The scenario of stretching for I1-1g1c at $\tilde{T}=0.3$.}
\end{figure}

\begin{figure}
\epsfxsize=3.2in
\centerline{\epsffile{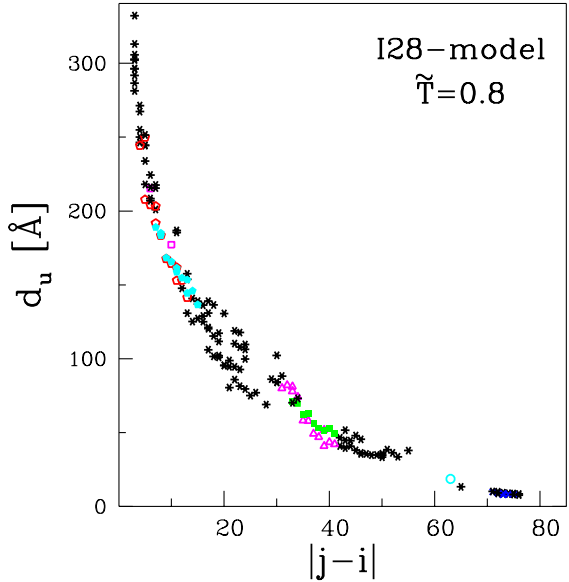}}
\caption{The scenario of stretching for I27-1tit at $\tilde{T}$=0.8.}
\label{fig:pastor8}
\end{figure}

\begin{figure}
\epsfxsize=3.2in
\centerline{\epsffile{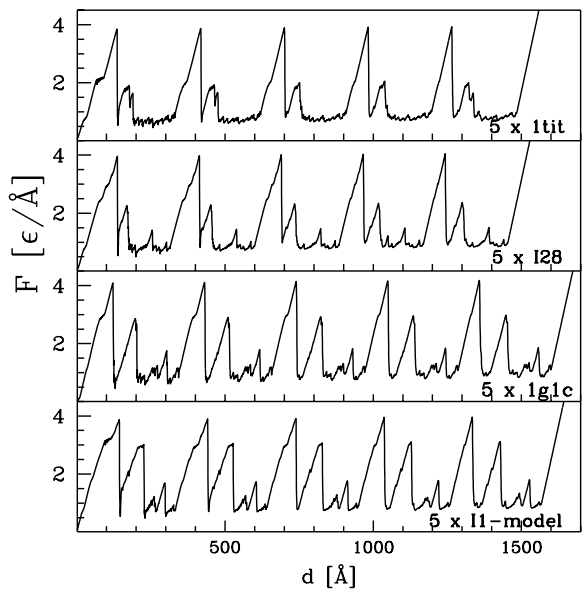}}
\caption{Stretching of five domains of titin, linked in a tandem
arrangement as listed in each panel, at $\tilde{T}$=0.0.}
\end{figure}

\noindent
{\bf Stretching of several domains of titin}\\

We now generate tandem arrangements of five identical domains of titin.
The $\tilde{T}$=0 $F-d$ patterns are shown in Figure 16.
To a good approximation, the patterns are a serial combination
of the single domain patterns of Figure 7: the proteins unwind
domain by domain.
The reason for this serial unwinding is that for a single domain
the largest force peak is located early in the unraveling process.
However, the larger the temperature, the more parallelism
in the unravelling\cite{Cieplak2004a}.
The small shoulder before the first peak points out that all domains
unfold simultaneously to a metastable state 
in which contacts between all terminal strands A and G 
and one contact (4-23) between all strands A and B break 
before the first domain unfolds completely. 
As we discussed above, this metastable state for I27 domain has been
also identified as the intermediate state \cite{Marszalek1999}.
Note that in Ref. \cite{Marszalek1999} the authors did not consider
the breaking of the A--G contacts as being an ingredient of the
intermediate state. We have found that a similar phenomenon takes 
place for the two versions of the I1 domain and for I28-model.
In all cases, the unfolding intermediate involves breaking of 
contacts between strands A and G and some contacts between strands
A and B. We observe that the extension at which this happens
is smaller in I1 and I28 than in I27. Figure 16 also shows
that the `hump' before the first peak is somewhat milder in the other
domains than in I27. Note that in Ref. \cite{Marszalek1999} the `hump'
is also observed for a tandem arrangement of I28.

On increasing the temperature, the simultaneous unwinding 
of the domains to the intermediate state still holds but also
thermal fluctuations gain in importance and affect the patterns 
significantly. Figure 17 shows the F--d patterns for various
tandems of domains at $\tilde{T}$ of 0.3. It is seen that
it is only the first segment that repeats
the single domain pattern fairly accurately
whereas the remaining segments in the
sawtooth-like pattern lose the last single domain maximum:
there is no second maximum in the fivefold repeat of I27-1tit
and no third maximum in the remaining fivefold repeats of
the other structures. The reason for this is that the second
maximum in I27-1tit and the third maximum in other structures
are barely stable. As unfolding of the domains continues, thermal
fluctuations (or kinetic energy) associated with the already
released chain length gain in strength
and destroy the peaks with low stability. Figure 17 shows that,
for tandem arrangements of I28-model (the second panel) and I1-1g1c 
(the third panel), the second peak becomes increasingly weak.
It actually disappears in the last segment of the sawtooth pattern
for the case of I28-model.

It can also be seen in the top three panels of Figure 17
that the heights of the maxima in the sawtooth patterns are 
the highest for the I1-1g1c domains.
Furthermore, the first four maxima for  the I28-model domains
are a bit weaker than those of I27-1tit even though they were
of about the same height for the single domains.
This indicates that the admixture of parallel unwinding that occurs
in tandem arrangements affect mechanical stability of the
individual domains.
In order to study the stability differences
of the domains better we have constructed 
heterogeneous tandem arrangements of the structures. In this case the
less stable domains will unfold before the more stable ones.
The two bottom panels in Figure 17 shows the $F-d$  curves for
two heterogeneous tandem arrangements of five domains at $\tilde{T}$=0.3.
Specifically, 3 domains of I27-1tit are linked, in 
an alternating fashion, with
two domains of I28-model or two domains of I1-1g1c.
We find that I1-1g1c is the domain that unravels the last, i.e.
is the most stable. I28-model, on the other hand, shows a more
complex behavior. We find that when placed in tandem with I27-1tit,
I28-model is the domain that unravels the first, but after the 
unraveling of the first domain, the two types of domain
can alternate -- the order of unfolding is determined by 
fluctuations. This indicates that, in tandem arrangements, I28-model 
is only slightly less stable
than I27 and the difference in mechanical stability 
between these two domains becomes insignificant comparing to thermal 
fluctuations after the first domain unfolds. This observation
is consistent with the experimental findings.\cite{Politou1996} 
As an example, the curve shown in the last panel of 
Figure 17, corresponds to unraveling proceeding in the order:
I28, I27, I27, I27, and I28.
Figure 18 shows the snapshots of the five-domain tandem arrangements
corresponding to $d$=500 $\AA$ and $\tilde{T}$=0.3 and illustrates
the domain-by-domain character of the unfolding in each case.

\begin{figure}
\epsfxsize=3.2in
\centerline{\epsffile{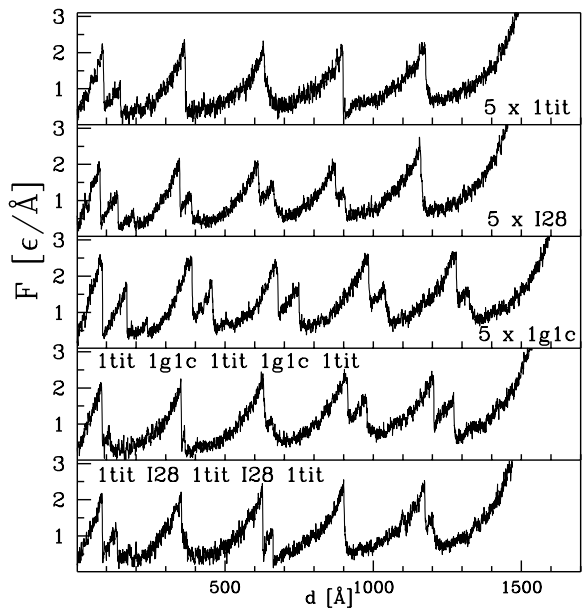}}
\caption{Stretching of five domains of titin, linked in a tandem
arrangement as listed in each panel, at $\tilde{T}$=0.3.}
\end{figure}

\begin{figure}
\epsfxsize=3.2in
\centerline{\epsffile{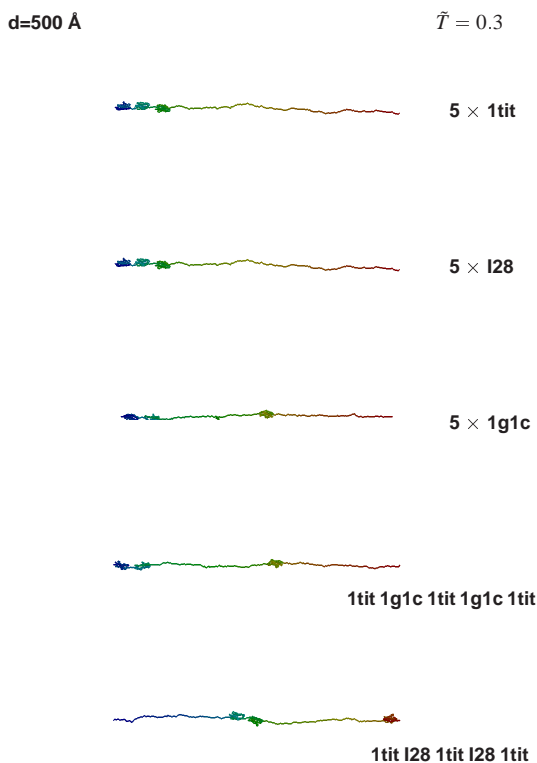}}
\caption{Snapshots of the five-domain arrangements when
stretched by 500 $\AA $. The top-to-bottom ordering of the snapshots
corresponds to the one adopted in Figure 11.}
\end{figure}

\begin{figure}
\epsfxsize=3.2in
\centerline{\epsffile{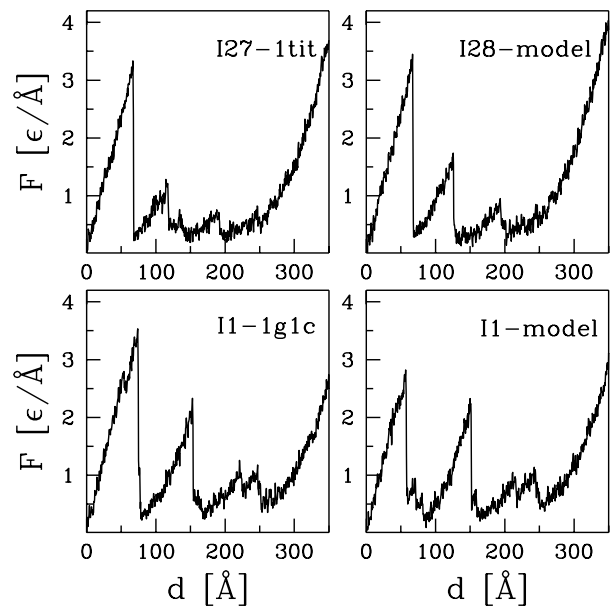}}
\caption{An analog of Figure 6 at $\tilde{T}$=0.3 for a generalized
model in which the hydrophobic-hydrophobic contacts are made
stronger by a factor of two.}
\end{figure}

\section{SUMMARY AND CONCLUSIONS}

In this paper, we have presented predictions of  Go-like modeling for 
four native structures of titin domains and of their tandem arrangements
and demonstrated that, despite noticeable differences in their native
structures, the force-displacement plots and the unravelling events
are qualitatively similar. Thus, the domains should be 
interchangeable in terms of their elastic properties. There are
also, however, substantial
differences when one considers finer details of the elastic
stability of the domains and their thermodynamic and kinetic properties
when the domains are considered in isolation.
I1-1g1c is the most stable structure in terms of its elastic resistance 
both separately and when placed in tandem arrangements. The I28-model 
domain 
has a mechanical stability close
to that of as single domain of I27, but it is clearly less stable 
in tandem arrangements that involve I27.

\begin{table*}
\caption{
Summary of the properties of the theoretically determined
structures corresponding to the I28 domain of titin as obtained
within the Go model used in this paper. The first column shows
a comparison to the similarly derived properties of I27-1tit. 
The lower the value of $T_f$, the lower the thermodynamic stability.
The values of the Z-score listed in the table are for all contacts.
They represent the deviations from the average quality value
as determined by using the program WHATIF with the NEWQUA option
in the QUALITY menu.
}
\begin{ruledtabular}
\begin{tabular}{lccccccc}
       & I27-1tit & I28-model & I28-A & I28-B & I28-C & I28-D &  I28-E \\
\hline
$F_{max}$ ($T=0$) [$\epsilon /\AA $]
     & 3.84 & 3.92 & 3.50 & 3.66 & 4.08 & 3.42 & 3.48 \\
$\tilde{T}_f$     & 0.185 & 0.224 & 0.174 & 0.172 & 0.192& 0.178 & 0.214 \\
$\tilde{T}_{min}$  & 0.275 & 0.30  & 0.25  & 0.275 & 0.25 & 0.275 &0.275 \\
$t_{fold}/\tau $ ($T=T_{min}$)& 3700  & 1730 & 2200 &
2040 & 1800 & 2480 & 1970  \\
$t_{fold}/\tau $ ($T=0.3$)  & 4320 & 1730 & 5440 &
2680 & 2600 & 2540 & 2060 \\
$F_{max}^{H}$ ($T=0$)\footnote{model with enhanced hydrophobic contacts} & 
4.92  & 5.06  & 4.86  & 4.30  &
4.89 & 4.32 & 4.72 \\
Z-score & -4.85 & -3.45 & -6.17  & -5.91 & -6.10 & -6.23 & -3.58 \\
\end{tabular}
\end{ruledtabular}
\end{table*}

More noticeable distinctions are observed for the thermodynamic and
kinetics properties of single domains.  We have shown that I1-1g1c and
I28-model are the two most thermodynamically stable structures and
I27-1tit is the least stable one. I28-model has a faster folding and
the range of temperature in which folding is optimal is wider than
I27; the same is observed for I1-model when compared to I1-1g1c. The
kinetics of refolding events
as studied as a function of the contact order
also show clear
distinctions. The contacts that form the last, on average, in the 
folding of I27-1tit are between strands C and F while for I28-model they 
are between strands B and E. 
In I1-1g1c and I1-model, the last events involve still
other sets of contacts. Only the earliest folding events are similar in
all models. All these differences show how sensitive the folding
properties of the models are to the precise details of the native
structures.

Most of our predictions about equivalence, or lack of it, of the
domains of titin in the context of the mechanical and kinetic
folding properties remains to be tested.
Currently, the I27 domain has been studied experimentally in the
most exhaustive way. The Go-like theoretical account
of mechanical properties of I27 has turned out to be consistent with
the experimental findings.\cite{Cieplak2002b,Cieplak2004b} 
Some 
experimental results available for the I28 domain, however, appear
to be at odds with our simulational results,
especially when one compares I28 to I27.
First of all, our studies predict
the stability of I28-model to be higher than that
of I27-1tit whereas the experiment shows the opposite:
the melting points for I27 and I28 are 72$^o$C and
35$^o$C respectively.\cite{Politou1995} 
Secondly, our folding simulations indicate
that I28-model folds faster than I27 whereas experiment shows that
the folding rate of an isolated I28 domain is 0.025 s$^{-1}$, 
which is
three orders of magnitude lower than that of I27 (32 s$^{-1}$)
\cite{Li2000} 
Finally, 
the atomic force microscopy studies\cite{Li2000} 
have shown that I28 domains are mechanically more
stable than I27 both in homo- and hetero-domain polyproteins
whereas our studies point to nearly equivalent stabilities.
This suggests that the method we are using cannot
easily describe cases such as I28 whose structure is very unstable in
solution. Any 'rigid' model which assumes a stable and compact structure
would therefore not be appropriate to represent the experimental 
conditions.

One source of the discrepancy
could be that our simple model incorporates just
one uniform energy scale $\epsilon$ whereas a more realistic
modeling would involve heterogeneous couplings.
This expected lack of homogeneity might govern the subtle differences
between the domains. In order to probe such effects, we have considered
a generalized Go-like model in which the interactions in contacts
that link two hydrophobic amino acids 
(ILE, LEU, MET, VAL, PHE, TRP, and TYR) are enhanced by the factor of two.
Figure 19
for $\tilde{T}=0.3$ shows that, mechanically, 
the relationships between the four domains 
in this two energy scale model are very much like as in 
the basic model. The noticeable difference though is that
I27-1tit acquires the more pronounced third peak in the force.
(The models with the two-fold enhancement of
the hydrophobic-hydrophobic contact strengths 
are inadequate kinetically: they give rise to an easy misfolding.)

Another source of the error then may be the
structure itself. The homology-based derivation of the structure may  
not sufficiently
refined for applications that involve dynamics.
Some hints can already be inferred by considering the I1 domain.
Our studies show that
I1-model and I1-1g1c, both meant to represent the same system,
have distinct kinetics and very different stabilities:
the  model structure has 259 contacts instead of 288 and is less
stable thermodynamically and mechanically
than the experimentally derived structure.
Even the force-displacement curves have distinct details indicating
that a precise knowledge of the structure may affect prediction of
dynamical properties in a substantial way.

We have thus generated five more homology-based structures of
the I28 domains, denoted as I28-A through I28-E. 
I28-model and I28-E are of the highest quality,
as judged from the standard
structure quality checks.\cite{Vriend1990}
The summary of their 
elastic, thermodynamic, and kinetic properties,
as determined within our Go-model, is given in Table I and compared
to those of I27-1tit. There are three conditions that should be
met for a model to agree with the experimental results:
1) I28 should be more stable mechanically than I27, but 2) it should
be less stable thermodynamically, and finally 3) it should refold
significantly slower. An inspection of Table I indicates that
none of the six I28 structures satisfies all three conditions.
At best, two conditions are met in some structures. For instance,
I28-C is the strongest mechanically of all of the I28 structures and 
it is also stronger than I27-1tit. Its folding temperature,
though not lower than that of I27-1tit, is
nearly to it. However, I28-C folds faster
both at $\tilde{T}_{min}$, when folding proceeds the fastest,
and at $\tilde{T}$=0.3 which appears to correspond to the room
temperature. 
Another good choice could be I28-A. 
This structure yields slower folding than I27-1tit at 
$\tilde{T}$=0.3 and is less stable thermodynamically. However,
its peak force is substantially smaller than that for I27-1tit.
Incompatibly with our predictions,
the Z-scores of both I28-C and I28-A are low which means that
these structures are not very good in terms of packing quality. The 
best Z-score among the new structures has I28-E, 
however this structure yields a mechanical
stability which is significantly lower than that of I27-1tit.

This analysis indicates that the precise definition of the structure 
has a substantial impact on the predicted properties. 
Existence of a well defined and fairly rigid native structure
is at the heart of applicability of the Go-like modeling.
In the context of the I28 domain, however, this feature may also be 
a root of the problem given that the experimentally determined
melting temperature of I28 is only 35$^o$ C. Nevertheless our
analysis illustrates a possibility that  using a dynamical
model may augment homology-based determination of protein structures.
The final message is that the simple Go-like description can capture
existence of differences in properties between various domains of titin
and of their various models.

{\bf Acknowledgments}

Support of M. C. from KBN in Poland  (grant number 2 P03B 025 13)
and from the European program IP NAPA through Warsaw University
of Technology is gratefully acknowledged.
T.~X.~H. thanks for support from the Natural Scientific
Council of Viet Nam.

\newpage


\vspace*{1cm}


\begin{thebibliography}{99} 

\bibitem{Carrion1999}
M. Carrion-Vasquez, P.~E. Marszalek, A.~F. Oberhauser, and J.~M. Fernandez,
Proc. Natl. Acad. Sci. USA {\bf 96}, 11288 (1999).

\bibitem{Carrion2000}
M. Carrion-Vazquez, A. F. Oberhauser, T. E. Fisher, P. E. Marszalek,
H. Li, and J. M. Fernandez,
Prog. Biophys.
Mol. Biol. {\bf 74}, 63 (2000).

\bibitem{Li2000}
H. Li, A.~F. Oberhauser, S.~B. Fowler, J. Clarke, and J.~M. Fernandez, 
Proc. Natl. Acad. Sci. USA {\bf 92}, 6527 (2000).

\bibitem{Maruyama1994}
K. Maruyama, 
Biophys. Chem. {\bf 50}, 73 (1994).

\bibitem{Maruyama1997}
K. Maruyama, 
FASEB J. {\bf 11}, 341 (1997).

\bibitem{Keller1995}
T.~C.~S. Keller,
Curr. Opin. Cell Biol. {\bf 7}, 32 (1995).

\bibitem{Trinick1996}
J. Trinick, 
Curr. Biol. {\bf 6}, 258 (1996).

\bibitem{Horowits1989}
R. Horowits, K. Maruyama, and R.~J. Podolsky,
J. Cell Biol. {\bf 109}, 2169 (1989).

\bibitem{Horowits1992}
R. Horowits, 
Biophys J. {\bf 61}, 392 (1992).

\bibitem{Funatsu1993}
T. Funatsu, et al., \& S. Tsukita,  
J. Cell Biol. {\bf 120}, 711 (1993).

\bibitem{Trombitas1991}
P.~H. Trombitas, W.~W. Baatsen, M.~S.~Z. Kellermayer and G.~H. Pollack 
J. Cell. Sci. {\bf 100}, 809 (1991).

\bibitem{Wang1991}
K. Wang, R. McCarter, J. Wright, J. Beverly, and R. Ramirez-Mitchell, 
Proc. Natl. Acad. Sci. USA {\bf 88}, 7101 (1991).

\bibitem{Labeit1995}
S. Labeit and B. Kolmerer, 
Science {\bf 270}, 293 (1995).

\bibitem{Linke1996}
W.~A. Linke, M. Ivemeyer, M. Olivieri, B. Kolmerer, C. Ruegg, and
S. Labeit, 
J. Mol. Biol. {\bf 261}, 62 (1996).

\bibitem{Linke1998}
W.~A. Linke, and H. Granzier,
Biophys. J. {\bf 75}, 2613 (1998).

\bibitem{Improta1996}
S. Improta, A. S. Politou, and A. Pastore
Structure {\bf 4}, 323 (1996).

\bibitem{Politou1995}
A.~S. Politou, D.~J. Thomas, and A. Pastore, 
Biophys. J. {\bf 69}, 2601 (1995).

\bibitem{Politou1996}
A.~S. Politou, M. Gautel, S. Improta, L. Vangelista, and A. Pastore,
J. Mol. Biol. {\bf 255}, 604 (1996).

\bibitem{Fowler2001}
S.~B. Fowler, and J. Clarke,  
Structure {\bf 9}, 355 (2001).

\bibitem{Carrion1999a}
M. Carrion-Vazquez, A.~F. Oberhauser, S.~B. Fowler, P.~E. Marszalek, S.~E.
Broedel, J. Clarke, and J.~M. Fernandez,
Proc. Natl Acad.  Sci. USA {\bf 96}, 3694 (1999).

\bibitem{Marszalek1999}
P.~E. Marszalek, H. Lu, H.~B. Li, M. Carrion-Vazquez, A.~F. Oberhauser,
K. Schulten, and J.~M. Fernandez,
Nature {\bf 402}, 100 (1999).

\bibitem{Best2001}
R.~B. Best, B. Li, A. Steward, V. Daggett, and J. Clarke, 
Biophys. J. {\bf 81}, 2344 (2001).

\bibitem{Lu1999}
H. Lu and K. Schulten,  
Chem. Phys. {\bf 247}, 141 (1999).

\bibitem{Lu2000}
H. Lu and K. Schulten, 
Biophys. J. {\bf 79}, 51 (2000).

\bibitem{Paci2000}
E. Paci and M. Karplus, 
Proc. Natl. Acad.  Sci. USA {\bf 97}, 6521 (2000).

\bibitem{Klimov2000}
D. K. Klimov and D. Thirumalai,  
Proc. Natl. Acad. Sci.  USA {\bf 97}, 7254 (2000).

\bibitem{Cieplak2002b}
M. Cieplak, T. X. Hoang, and M. O. Robbins, 
Proteins {\bf 49}, 114 (2002).

\bibitem{Cieplak2004a}
M. Cieplak, T. X. Hoang, and M. O. Robbins,
Phys. Rev. E {\bf 69}, 011912 (2004).

\bibitem{Cieplak2004b}
M. Cieplak, T. X. Hoang,  and M. O. Robbins, 
Proteins: Struct. Funct. Bio. {\bf 56}, 285 (2004).

\bibitem{Abe1981}
H. Abe and N. Go, 
Biopolymers {\bf 20}, 1013 (1981).

\bibitem{Takada1999}
S. Takada 
Proc. Natl. Acad. Sci. USA {\bf 96}, 11698 (1999).

\bibitem{Fraternali1999}
F. Fraternali and A. Pastore, 
J. Mol. Biol. {\bf 290}, 581 (1999).

\bibitem{Mayans2001}
O. Mayans, J. Wuerges, M. Gautel, and M. Wilmans,
Structure {\bf 9}, 331 (2001).

\bibitem{Bernstein1997}
F. C. Bernstein, T. F. Koetzle, G. J. B. Williams, E. F. Meyer Jr.,
M. D. Brice, J. R. Rodgers, O. Kennard, T. Shimanouchi, and M. Tasumi,
J. Mol. Biol. {\bf 112}, 535 (1997).

\bibitem{Sali1993}
A. Sali, T. L. Blundell, 
J. Mol. Biol. {\bf 234}, 779 (1993).

\bibitem{Vriend1990}
G. Vriend, 
J. Mol. Graph.  {\bf 8}, 52 (1990).

\bibitem{Guex1997}
N. Guex and M. C. Peitsch 
Electrophoresis {\bf 18}, 2714 (1997).

\bibitem{Thompson1997}
J. D. Thompson, T. J. Gibson, F. Plewniak, F. Jeanmougin, and 
D. G. Higgins,
Nucleic Acids Res. {\bf 25}, 4876 (1997).

\bibitem{Hoang2000}
T. X. Hoang,  and M. Cieplak, 
J. Chem. Phys. {\bf 112}, 6851 (2000).

\bibitem{Hoang2001}
T. X. Hoang, and M. Cieplak, 
J. Chem. Phys. {\bf 113}, 8319 (2001).

\bibitem{Cieplak2003a}
M. Cieplak, and T. X. Hoang, 
Biophys. J. {\bf 84}, 475  (2003).

\bibitem{Cieplak2002}
M. Cieplak and T. X. Hoang, 
Int. J. Mod. Phys. C {\bf 13}, 1231 (2002).

\bibitem{Tsai1999}
J. Tsai, R. Taylor, C. Chothia, and M. Gerstein, 
J. Mol. Biol. {\bf 290}, 253 (1999).

\bibitem{Cieplak2003b}
M. Cieplak and T. X. Hoang, 
Physica A {\bf 330}, 195 (2003).

\bibitem{Kwiecinska}
J. I. Kwiecinska and M. Cieplak,
J. Phys. Cond. Mat. (submitted).

\bibitem{Veitshans1997}
T. Veitshans , D. Klimov, and D. Thirumalai, 
Folding Des. {\bf 2}, 1 (1997).

\bibitem{Cieplak2002a}
M. Cieplak, T. X. Hoang, and M. O. Robbins, 
Proteins {\bf 49}, 104 (2002).

\bibitem{Rief1997}
M. Rief, M. Gautel, F. Oesterhelt, J. M. Fernandez,  and  H. E. Gaub, 
Science {\bf 276}, 1109 (1997).


\bibitem{Fowler2002}
S. B. Fowler, R. B. Best, J. L. Toca Herrera, T. J. Rutherford, A. Steward,
E. Paci, M. Karplus M, and J. Clarke, 
J. Mol. Biol. {\bf 322}, 841 (2002).

\bibitem{Marko1995}
J. F. Marko. and E. D. Siggia, 
Macromol. {\bf 28}, 8759 (1995)

\end{thebibliography}
\end{document}